\documentstyle[aps,prd,epsfig,floats]{revtex} 

\draft

\begin{document}
\twocolumn[\hsize\textwidth\columnwidth\hsize\csname
@twocolumnfalse\endcsname
\title{%
\hbox to\hsize{\normalsize\rm \hfil Preprint MPI-PTh/98-28}
\vskip 32pt
The Bremsstrahlung Process 
{\boldmath$\nu_\tau\to\nu_e e^+ e^- \gamma$}}
\author{H.~Schmid and G.~Raffelt}
\address{Max-Planck-Institut f\"ur Physik 
(Werner-Heisenberg-Institut),
F\"ohringer Ring 6, 80805 M\"unchen, Germany}
\author{A.~Leike}
\address{Ludwig-Maximilians-Universit\"at, Sektion
Physik, Theresienstr.~37, 80333 M\"unchen, Germany}
\date{April 20, 1998}
\maketitle
\begin{abstract}
The bremsstrahlung process $\nu_\tau\to\nu_e e^+e^-\gamma$ is the
dominant photon-producing decay of heavy $\tau$ neutrinos with
$m_{\nu_\tau}>2m_e$ so that this process is instrumental for supernova
1987A constraints on the properties of these particles, notably on
their mixing amplitude with $\nu_e$. We calculate the rate of the
bremsstrahlung process as well as the photon spectrum and the angular
distribution relative to the spin direction of the parent neutrino. We
carefully pay attention to the difference between Dirac and Majorana
neutrinos.
\end{abstract}
\vskip2.0pc]


\section{Introduction}

The direct experimental limits on neutrino masses are so crude that
astrophysical arguments remain the most important source of
information on their possible magnitude. The well-known cosmological
limit~\cite{KolbTurner} $m_\nu\alt40~\rm eV$ improves the experimental
$m_{\nu_\tau}$ limit of about $18.2~\rm MeV$~\cite{NuTauMass} by
almost six orders of magnitude, assuming that it is stable on
cosmological time scales. This longevity cannot be taken for granted
because the standard-model decay $\nu_\tau\to \nu_e e^+ e^-$ is fast
on cosmic time scales even for relatively small mixing angles between
$\nu_\tau$ and $\nu_e$.  Therefore, to plug this loophole on the
cosmological $m_{\nu_\tau}$ limit, one needs restrictive limits on the
$\nu_e$-$\nu_\tau$ mixing angle. Even then, of course, heavy $\tau$
neutrinos could decay fast by new interactions so that the
cosmological mass bound is never absolute.  In fact, heavy
$\nu_\tau$'s with ``invisible'' fast decays can fix certain problems
of the standard cold dark matter cosmology, leading to the interesting
class of cosmological $\tau$CDM models~\cite{tauCDM}.  In any case,
proving that a heavy $\nu_\tau$ requires interactions beyond the
particle-physics standard model to escape the cosmological mass limit
is a significant constraint on its possible properties.

If $\nu_e$ and $\nu_\tau$ mix with each other, any $\nu_e$ source also
produces a certain $\nu_\tau$ flux. Experimental limits on the flux of
decay electrons and positrons from beam-dump~\cite{Leener},
reactor~\cite{Hagner}, and solar neutrinos~\cite{Toussaint} provide
strong limits on the mixing amplitude. The most restrictive limit
arises from core-collapse supernovae which are powerful $\nu_\tau$
sources because they produce this flavor directly rather than by an
assumed mixing with $\nu_e$.  The decay positrons would linger in the
galaxy for about $10^5$ years until they annihilate so that one can
derive restrictive limits on the $\nu_\tau\to\nu_e e^+ e^-$ channel
from the observed interstellar positron flux
\cite{Dar87a,Zhang,RaffeltBook}. In addition, the absence of
$\gamma$-rays associated with the neutrino burst from supernova (SN)
1987A provides further limits where the bremsstrahlung process
$\nu_\tau\to \nu_e e^+ e^-\gamma$ is the dominant photon-producing
reaction~\cite{Zhang,RaffeltBook,Dar87b,Jaffe}.

This literature has two important gaps. First, no detailed calculation
of $\nu_\tau\to \nu_e e^+ e^-\gamma$ is available.  While the simple
estimates used in Refs.~\cite{Zhang,RaffeltBook,Dar87b,Jaffe} are
certainly adequate for a first estimate, a detailed calculation of the
expected $\gamma$-ray spectrum from a supernova depends on the angular
distribution and spectrum of the photons in a given $\nu_\tau\to \nu_e
e^+ e^-\gamma$ decay.  It is the purpose of our paper to provide this
missing calculation.

A second gap concerns the data used for the derivation of the limits.
Refs.~\cite{Zhang,RaffeltBook,Dar87b} used $\gamma$-ray data that had
been taken by the Gamma Ray Spectrometer on the Solar Maximum Mission
satellite in coincidence with the SN~1987A neutrino signal.
Ref.~\cite{Jaffe} used $\gamma$-ray data from the Pioneer Venus
Orbiter spacecraft that were also taken in conicidence with the
SN~1987A burst. However, to constrain radiative decays from MeV-mass
neutrino decays it is not critical to use data even close to the
$\bar\nu_e$ burst because the expected $\gamma$ pulse is widely
dispersed. The COMPTEL instrument aboard the Compton Gamma Ray
Observatory satellite looked at the SN 1987A remnant for about
$0.68\times10^6~\rm s$ in 1991~\cite{Miller}, much longer than the
previous experiments.  Because of the long COMPTEL viewing period, one
could derive limits on the $\nu\to\nu'\gamma$ channel which are far
superior to the SMM and PVO results, assuming that the neutrino mass
exceeds about $0.1~\rm MeV$~\cite{Miller}. For smaller masses the
$\gamma$ burst would have ended before COMPTEL looked at the SN so
that the previous limits remain of interest. In the $\nu_\tau\to \nu_e
e^+e^-$ scenario it is necessarily assumed that $m_{\nu_\tau}\agt
1~\rm MeV$ so that the COMPTEL limits are naturally the most
restrictive. However, the authors of Ref.~\cite{Miller} have only
analysed the direct channel $\nu_\tau\to\nu_x\gamma$. Our
bremsstrahlung calculation could be used to extend the interpretation
of the COMPTEL data to the more interesting $\nu_\tau\to \nu_e
e^+e^-\gamma$ case.

Aside from the practical application in the supernova context, the
bremsstrahlung process turns out to be another interesting example for
the differences between massive Dirac and Majorana neutrinos.

We proceed in Sec.~II with a brief discussion of the bare process
$\nu_\tau\to\nu_e e^+ e^-$. In Sec.~III we calculate the rate,
spectrum, and angular distribution of the bremsstrahlung process
$\nu_\tau\to\nu_e e^+ e^-\gamma$.  In Sec.~IV we discuss and summarize
our results.


\section{The Bare Process}

We begin with a discussion of the bare decay process
$\nu_3\to\nu_1e^+e^-$ where we assume a neutrino mass hierarchy of the
form $m_3>m_2>m_1$. The $m_3$ eigenstate is taken to be the dominant
admixture of the $\nu_\tau$ flavor eigenstate, and similarly $m_1$ is
the dominant mass component of the electron neutrino. This process is
then essentially what we sloppily called $\nu_\tau\to\nu_ee^+e^-$ in
the introduction.

\begin{figure}[b]
\center\leavevmode
\epsfig{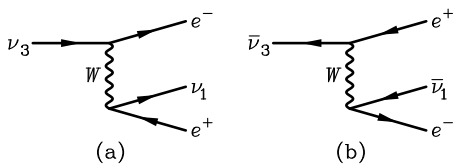}
\smallskip
\caption{\label{fig:graphs}
Decay amplitude for a heavy neutrino.} 
\end{figure}

\begin{figure}[ht]
\center\leavevmode
\epsfig{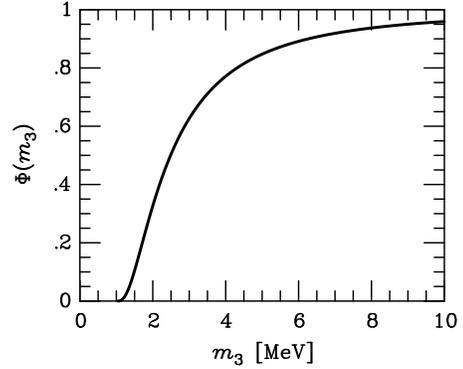}
\smallskip
\caption{\label{fig:phasespace}
Phase-space factor according to Eq.~(\protect\ref{eq:phasespace}).} 
\end{figure}

For a Dirac neutrino the relevant amplitude is the tree-level graph of
Fig.~\ref{fig:graphs}(a) which was evaluated by Shrock~\cite{Shrock}.
The decay rate is found to be
\begin{eqnarray}
\Gamma_{\rm bare}&=&
\vert U_{e3}U_{e1}\vert^2\,\frac{G_F^2}{3\,(4\pi)^3}\,m_3^5\,\Phi(m_3)
\nonumber\\
\noalign{\vskip4pt}
&=&\vert U_{e3}U_{e1}\vert^2
\,3.5{\times}10^{-5}\,{\rm s}^{-1}
\,\left(\frac{m_3}{\rm 1\,MeV}\right)^5\,\Phi(m_3),
\end{eqnarray}
where $U_{e1}$ and $U_{e3}$ are the mixing amplitudes between $\nu_e$
and $\nu_1$ or $\nu_3$, respectively, and $G_F$ is Fermi's constant.
It was assumed that $m_1$ is negligibly small.  The phase-space factor
(Fig.~\ref{fig:phasespace}) is 
\begin{eqnarray}\label{eq:phasespace}
\Phi(m_3)&=&(1-4a)^{1/2}\,(1-14a-2a^2-12a^3)\nonumber\\
&+&24a^2(1-a^2)\ln\frac{1+(1-4a)^{1/2}}{1-(1-4a)^{1/2}}
\end{eqnarray}
with $a\equiv (m_e/m_3)^2$. $\Phi$ approaches unity for $m_3\gg m_e$.

This process was also studied by Li and Wilczek~\cite{Li} with an
emphasis on the difference between Dirac and Majorana
neutrinos. Following their treatment we note that for Majorana
neutrinos the decay rate is effectively the incoherent sum of the two
Dirac graphs depicted in Figs.~\ref{fig:graphs}(a) and (b). While Li
and Wilczek focussed on the angular distribution of the final-state
charged leptons and did not explicitly compare the absolute rates,
their treatment implies that the Majorana decay rate is twice the
Dirac one, a result which we also find by an independent evaluation of
the relevant S-matrix element. The phase-space factor of
Eq.~(\ref{eq:phasespace}) is the same in both~cases.

\begin{figure}[b]
\center\leavevmode
\epsfig{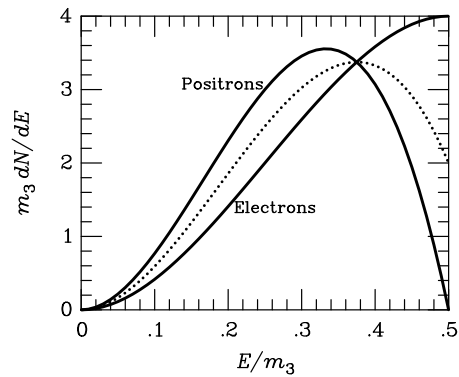}
\smallskip
\caption{\label{fig:barespectra} Positron and electron spectrum for
$\nu_3\to\nu_1e^+e^-$ according to Eq.~(\protect\ref{eq:barespectra})
for a Dirac neutrino with $m_3\gg m_e$.  The dotted line is the
average between the two spectra which pertains to both $e^+$ and $e^-$
when $\nu_3$ is a Majorana particle.}
\end{figure}

If $\nu_3$ is a Dirac neutrino the normalized positron and electron
spectra $dN/dE_\pm=\Gamma_{\rm bare}^{-1}\,d\Gamma_{\rm bare}/dE_\pm$
are
\begin{eqnarray}\label{eq:barespectra}
\frac{dN}{dE_+}&=&\Theta (q^{2}-m_{e}^{2})\,\frac{96}{m_{3}^{5}}\,
\frac{(E_{+}^{2}-m_{e}^{2})^{1/2}}{\Phi (m_{3})}\,
f(q^{2})\,q^2 E_+,\nonumber\\
\frac{dN}{dE_-}&=&\Theta(q^{2}-m_{e}^{2})\,
\frac{32}{m_{3}^{5}}\,\frac{(E_{-}^{2}-m_{e}^{2})^{1/2}}
{\Phi (m_{3})}\,f(q^2)\nonumber\\
&\times&
\biggl[\frac{q^2-m_e^2}{2}\,E_{-}\nonumber\\
&+&\frac{q^2+2m_e^2}{q^2}\,(m_{3}-E_{-})\,(m_{3}E_{-}-m_{e}^{2})\biggr],
\end{eqnarray}
where $q^{2}=m_{3}^{2}-2m_{3}E_{\pm}+m_{e}^{2}$ and
$f(q^{2})=(1-m_{e}^{2}/q^{2})^2$.  For a parent neutrino mass $m_3\gg
m_e$ these spectra are shown in~Fig.~\ref{fig:barespectra}.  For
$\bar\nu_3$ the role of electrons and positrons is interchanged.  For
a Majorana parent the electron and positron spectra are the same. Each
is the average of the Dirac spectra of 
Eq.~(\ref{eq:barespectra})---see the dotted line in
Fig.~\ref{fig:barespectra}. 


\section{Bremsstrahlung}

\subsection{Matrix Element}

For a Dirac neutrino the amplitude for $\nu_3\to\nu_1e^+e^-\gamma$ is
given by the Feynman graph of Fig.~\ref{fig:graphs}(a) with a photon
attached to either the electron or positron line. We calculate the
squared matrix element, summed over all final-state spin and
polarization states.  However, because relativistic Dirac neutrinos
are produced primarily with negative helicities we do not average over
the initial-state helicities---a helicity-averaged rate would not
pertain to a realistic source. We find
\begin{eqnarray}\label{eq:matrixelement}
&&\sum_{\rm final-state~spins \atop polarizations}|{\cal M}|^{2}=
\left(\frac{e\,G_F}{\sqrt{2}}\right)^2 |U_{e1}U_{e3}|^2
\nonumber\\
&&{}\times\frac{8}{C_{\gamma-}^2}
\Bigl(C_{\gamma-}C_{3+}C_{1\gamma}
-C_{1-}C_{3+}m_e^2
-C_{3+}C_{1\gamma}m_e^2\Bigr)\nonumber\\
&&{}+\frac{8}{C_{\gamma+}^2}
\Bigl(C_{1-}C_{\gamma+}C_{3\gamma}-C_{1-}C_{3+}m_e^2
-C_{1-}C_{3\gamma}m_e^2\Bigr)\nonumber\\
&&{}+\frac{8}{C_{\gamma-}C_{\gamma+}}
\Bigl(2C_{-+}C_{1-}C_{3+}+C_{-+}C_{1-}C_{3\gamma}\nonumber\\
&&\hskip3em{}+C_{-+}C_{3+}C_{1\gamma}-C_{3-}C_{1-}C_{\gamma+}
+C_{1-}C_{\gamma-}C_{3+}\nonumber\\
&&\hskip3em{}+C_{1-}C_{3+}C_{\gamma+}-C_{\gamma-}C_{3+}C_{1+}\Bigr),
\end{eqnarray}
where $C_{1+}\equiv P'_1P'_+$ and so forth with
$P'_{1,+,-,\gamma}=P_{1,+,-,\gamma}$ the four momenta of $\nu_1$,
$e^+$, $e^-$, and $\gamma$, respectively.  Only for the parent
neutrino we have $P'_3=P_3-m_3 S_3$ with $S_3$ its spin four vector;
we have not summed over its helicities. In the $\nu_3$ rest
frame $P_3'=m_3(1,-{\bf s}_3)$ with ${\bf s}_3$ a unit
vector in the spin direction of $\nu_3$.

If the initial state is a Dirac antineutrino we find the same squared
matrix element with the role of $e^+$ and $e^-$ interchanged.
Moreover, we then have $P'_3=P_3+m_3 S_3$, leading to
$P_3'=m_3(1,+{\bf s}_3)$ in the $\bar\nu_3$ rest frame. A simple CP
transformation reveals that the photon angular distribution relative
to the spin-polarization vector is opposite compared to that of a
neutrino. Put another way, neutrinos and antineutrinos with the same
momentum but opposite helicities will produce identical angular photon
distributions relative to the momentum direction.

For Majorana neutrinos the squared matrix element is the sum of
Eq.~(\ref{eq:matrixelement}) with the same expression under the
exchange of $e^+$ with $e^-$. In this case the photon angular
distribution in the parent rest frame is isotropic, independently of
its helicity state. For Majorana neutrinos, then, one may average
over the initial-state neutrino helicities (one may use $P_3'=P_3$)
without loss of generality.

\subsection{Photon Spectrum}

In order to calculate the photon spectrum we need to integrate over
the four-body final-state phase space, leaving only the $d\omega$
integration undone, where $\omega$ is the photon energy. The shape of
the photon spectrum, integrated over all emission angles, is the same
for Dirac neutrinos or antineutrinos as well as for Majorana
neutrinos, independently of their polarization state.  We have
performed the phase-space integration numerically by a Monte Carlo
technique with the parametrization outlined in
Appendix~\ref{app:phase-space}.

To derive a useful representation of the resulting spectrum we borrow
from a standard result for the classical spectrum of an inner
bremsstrahlung process~\cite{Jackson}
\begin{equation}\label{eq:classical}
\frac{d\Gamma}{d\omega}=\frac{\Gamma_{\rm bare}}{\omega}
\,\frac{\alpha}{\pi}\,
\left[\frac{1}{v}\,\ln\left(\frac{1+v}{1-v}\right)-2\right],
\end{equation}
where $\omega$ is the photon energy, $\Gamma_{\rm bare}$ the rate for
the bare process, $\alpha\approx1/137$ the fine-structure constant,
and $v$ the speed of the produced charged lepton. This calculation
applies to a process like beta decay where the recoiling nucleus does
not contribute to the bremsstrahlung rate. In our case we have two
charged leptons so that we expect twice this rate, for the moment
ignoring interference effects. Of course, the $1/\omega$ divergence
represents the well-known infra-red bremsstrahlung behavior.

Motivated by this classical result we represent the photon spectrum by
\begin{equation}\label{eq:spectrum}
\frac{d\Gamma}{d\omega}=\frac{\Gamma_{\rm bare}}{\omega}
\,\frac{2\alpha}{\pi}\,A\,g(x),
\end{equation}
where $g(x)$ is a dimensionless function which depends on the parent
neutrino mass $m_3$. The dimensionless coefficient $A$ is factored out
such that $g(0)=1$, i.e.\ $g(x)$ describes the spectral shape, not its
absolute amplitude. The factor $\Gamma_{\rm bare}$ includes a global
factor of 2 for Majorana neutrinos as discussed in section~II.  The
dimensionless photon energy $x\equiv\omega/\omega_{\rm max}$ is in the
range $0\leq x\leq1$, where
\begin{equation}
\omega_{\rm max}=\frac{m_3}{2}\,
\biggl[1-\left(\frac{2m_e}{m_3}\right)^2\biggr]
\end{equation}
is the end point of the photon spectrum. 

\begin{figure}[t]
\center\leavevmode
\epsfig{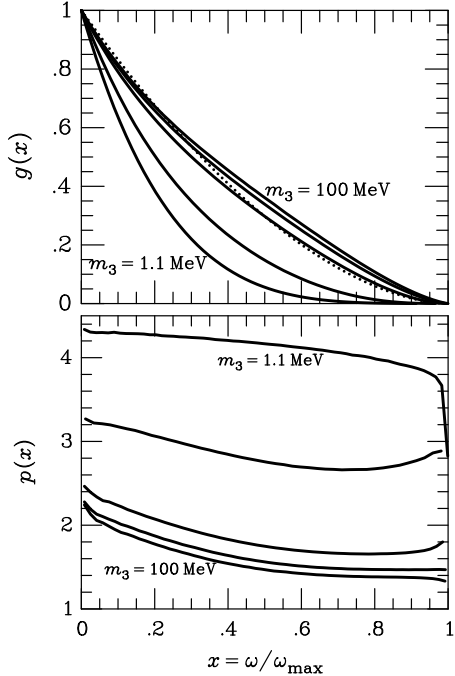}
\smallskip
\caption{\label{fig:spectrum} {\em Upper Panel:} Numerical results for
the dimensionless bremsstrahlung photon spectrum $g(x)$ as defined in
Eq.~(\protect\ref{eq:spectrum}) for parent neutrino masses $m_3=1.1$,
2, 10, 30, and 100~MeV (bottom to top).  The dotted line is a power
law of the form Eq.~(\protect\ref{eq:powerlaw}) with $p=1.75$.  {\em
Lower Panel:} Power-law index $p(x)$ as defined in
Eq.~(\protect\ref{eq:powerlaw}) for the same parent masses (top to
bottom).}
\end{figure}

In the upper panel of Fig.~\ref{fig:spectrum} we show $g(x)$ for
several values of $m_3$. For all cases $g(x)$ decreases monotonically
from 1 at $x=0$ to 0 at $x=1$. This behavior suggests a 
power-law representation
\begin{equation}\label{eq:powerlaw}
g(x)=(1-x)^{p(x)},
\end{equation}
where the power-law index $p(x)$ is itself a slowly varying function
of $x$. We show it in the lower panel of Fig.~\ref{fig:spectrum} for
several values of $m_3$. In a practical application it would probably
be enough to use a constant, typical value for $p$. For reference we
show in the upper panel as a dotted line $g(x)$ for a fixed value
$p=1.75$ which would appear to produce a reasonably good
approximation if $m_3$ is taken to be around 10~MeV.

The dimensionless coefficient $A$ measures the zero-point intercept of
the spectrum. Numerical values for several parent masses are given in
Table~\ref{tab:a-values}. 

\begin{table}[b]
\caption{\label{tab:a-values}
Values for the coefficient $A$ as defined in
Eq.~(\protect\ref{eq:spectrum}).}
\smallskip
\begin{tabular}{rll}
$m_3$ [MeV]&\multicolumn{1}{c}{$A$ (quantum)}&
\multicolumn{1}{c}{$A$ (classical)}\cr
\noalign{\vskip 2pt}
\hline
\noalign{\vskip 2pt}
100.0&7.99 &7.55   \cr
 30.0&5.58 &5.16   \cr
 10.0&3.44 &3.06   \cr
  2.0&0.76 &0.54   \cr
  1.1&0.068&0.036  \cr
\end{tabular}
\end{table}

It is a useful check of our calculation to compare these numerical
values with the classical result Eq.~(\ref{eq:classical}) which should
be exact in the limit of soft photons. However, it does not include
the interference between the radiation emitted by both final-state
charged leptons. What we list in Table~\ref{tab:a-values} as the
classical prediction, then, is the incoherent sum of the expected
radiation power from the final-state electron and positron where we
have integrated over the velocity distribution of these particles as
given by the bare process in Eq.~(\ref{eq:barespectra}). This
classical result is always smaller, indicating that the interference
effect is constructive on average. In any case, for high parent masses
the importance of interference disappears as one would have expected
because then the bremsstrahlung photons are primarily emitted co-linear
with the charged leptons so that the overlap of the two
``bremsstrahlung cones'' becomes small. The most important conclusion
is that we do not seem to  have lost factors of 2, $\pi$, and so forth
in the course of our calculation. 

\subsection{Angular Distribution}

The photon angular distribution is isotropic for Majorana neutrinos
while it is nontrivial in the Dirac case. A possible angular
distribution can arise only from the terms
$C_{3\gamma}=(P_3-m_3S_3)P_\gamma$ in the squared matrix element
Eq.~(\ref{eq:matrixelement}). Because it is linear in $C_{3\gamma}$ it
is clear that after all unobserved phase-space degrees of freedom have
been integrated out we are left with a term which is independent of
the angle $\theta$ between the photon momentum and the spin-direction,
and one that is proportional to $\cos\theta$.  (The signs are chosen
such that $\cos\theta=+1$ corresponds to the direction along the
parent spin.)  Thus for a fixed photon energy $\omega$ the most
general normalized angular distribution is of the form
\begin{equation}\label{eq:angular}
\frac{dN_\gamma}{d\cos\theta}=\frac{1+a \cos\theta}{2}
\end{equation}
where $-1\leq a\leq+1$. Again, because $\theta$ is measured relative
to the spin direction, we have $a\to -a$ for antineutrinos.  For
Majorana neutrinos we have $a=0$ for all $\omega$ as mentioned above.

\begin{figure}[b]
\center\leavevmode
\epsfig{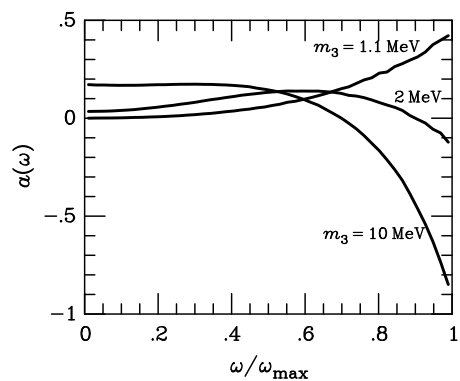}
\smallskip
\caption{\label{fig:angular}
Photon angular-distribution parameter $a(x)$ as defined in
Eq.~(\protect\ref{eq:angular}) for several choices of parent neutrino
mass.}
\end{figure}

In Fig.~\ref{fig:angular} we show $a$ as a function of the
dimensionless photon energy $x=\omega/\omega_{\rm max}$ for several
choices of parent neutrino mass. Contrary to what one might have
expected the angular distribution depends very sensitively on the
photon energy. Soft photons are emitted nearly isotropically, with a
slight bias in the spin direction. The highest-energy photons
are emitted dominantly against the spin direction if $m_3\gg m_e$.  In
the limit $m_3\to\infty$ it is easy to show analytically that $a=-1$
for $x=1$.  (The reverse for antineutrinos.)


\section{Discussion and Summary}

We have calculated the rate, photon spectrum, and photon angular
distribution of the inner bremsstrahlung process $\nu_\tau\to\nu_e
e^+e^-\gamma$ which could have produced a $\gamma$ ray burst from
SN~1987A, contrary to several observations. Therefore, one can derive
limits on the relevant neutrino mixing angle even though the most
relevant data, the 1991 COMPTEL observations of SN~1987A
\cite{Miller}, have not yet been used in this regard.

After assuming a certain magnitude and spectrum of the primary
$\nu_\tau$ and $\bar\nu_\tau$ flux, a prediction of the bremsstrahlung
photon flux and spectrum depends on the photon energy distribution as
well as their angular distribution relative to the neutrino
momentum. 

The simplest case is that of Majorana neutrinos where the angular
distribution in the neutrino rest frame is isotropic, independently of
the parent polarization. The rough estimate of the bremsstrahlung rate
that had been used in previous papers is
$d\Gamma/d\omega=(\alpha/\pi)\,\Gamma_{\rm bare}/\omega$ with photon
energies $\omega$ up to half the parent neutrino mass. It was
overlooked that for Majorana neutrinos $\Gamma_{\rm bare}$ is twice
that of Dirac neutrinos. We have a further factor 2 because
bremsstrahlung arises from two final-state charged leptons. Finally,
one gains the factor $A$ which is given in Table~\ref{tab:a-values}.
For example, taking $m_{\nu_\tau}=10~\rm MeV$ one gains overall more
than a factor of 10. However, the spectrum decreases somewhat faster
with $\omega$ than had been assumed.

Therefore, the previous limits on the mixing angle will improve
accordingly. They would improve much further by applying our results
to the COMPTEL data which encompass a much larger viewing period than
the previously used SMM and PVO data.  It would an important future
project to analyse the COMPTEL data with regard to the $\nu_\tau\to
\nu_e e^+ e^- \gamma$ decay!

Dirac neutrinos exhibit a nontrivial photon angular distribution if
they are polarized. Because the $\nu_\tau$'s and $\bar\nu_\tau$'s
emitted from a SN have opposite polarizations, the anisotropy of the
photon distribution does not cancel between them. On the other hand,
for a mass in the $10~\rm MeV$ range and typical energies of around
$30\,\rm MeV$ they are not particularly relativistic so that their
degree of polarization is incomplete. For smaller masses the degree of
polarization is stronger, but the deviation from isotropy is less
pronounced.  A detailed treatment of the Dirac case is probably too
complicated to be worth conducting in view of the many uncertainties
with regard to the overall neutrino flux and spectrum.

We do not believe that interpreting, for example, the COMPTEL data
requires an implementation of our bremsstrahlung results in every
detail. (One would need numerical results on a finely spaced grid of
parameters.)  Rather, we think that our results should be used as a
basis for an approximate treatment where the magnitude of the possible
errors is understood and controlled.

Independently from the SN application, the bremsstrahlung
process features differences between Dirac and Majorana neutrinos
which are interesting in their own right.


\section*{Acknowledgments}

This work is based on a thesis submitted by H.S.\ to the
Ludwig-Maximilians-Universit\"at, M\"unchen, in partial fulfillment of
the requirements for a {\it Diplom\/} degree (Master of Science).
Partial support by the Deutsche For\-schungs\-ge\-mein\-schaft under
grant No.\ SFB~375 is acknowledged.


\appendix

\section{Phase-Space Integration}
\label{app:phase-space}

We give some details about the integration of the four-particle phase
space in the $\nu_3\to\nu_1 e^+ e^-\gamma$ process.
Following Ref.~\cite{Bardin} a four-body phase-space integral can be
transformed according to
\begin{eqnarray}
&&\int \prod_{i=1}^4 \frac{d^3 {\bf p}_i}{2E_i}
\delta^4\left(P_{\rm ini}-\sum_{i=1}^4P_i \right)\nonumber\\
&&=\frac{\pi}{256}\int
\frac{\sqrt{\lambda(s,s_a,s_b)}}{s}\,
\frac{\sqrt{\lambda(s_a,m^2_1,m^2_2)}}{s_a}\nonumber\\
&&\hskip4em\times
\frac{\sqrt{\lambda(s_b,m^2_3,m^2_4)}}{s_b}\,ds_a ds_b\, d\cos\theta\,
d\Omega_a d\Omega_b,
\end{eqnarray}
with the four momenta $P_i=(E_i,{\bf p}_i)$, $i=1,\ldots,4$, and the
masses $m_i^2=P_i^2$. Further, we use
\begin{equation}
\lambda(a,b,c)=a^2+b^2+c^2-2ab-2ac-2bc.
\end{equation}
The four particles have been grouped into two subsystems with
$s_a=(P_1+P_2)^2$ and $s_b=(P_3+P_4)^2$ while
$s^2=(P_1+P_2+P_3+P_4)^2=m_{\rm ini}^2$. The angles $\Omega_{a,b}$ are
within the CM frames of the two subsystems while $\theta$ is the angle
with which the two subsystems move in opposite directions relative to
some chosen direction in the overall CM frame, i.e.\ in the rest frame
of the decaying particle, for example relative to its spin. The limits
of integration are
\begin{eqnarray}
&&(m_1+m_2)^2\leq s_a\leq (\sqrt{s}-m_3-m_4)^2,\nonumber\\
&&(m_3+m_4)^2\leq s_b\leq (\sqrt{s}-\sqrt{s_1})^2.
\end{eqnarray}
Finally we need to express the original four momenta $P_i$
in terms of the
new integration variables. 
Assuming that the two subsystems move in opposite directions along
the $z$-axis we find in the CM frame
\begin{equation}
P_1=\pmatrix{\gamma^{0}_{a}E^{R}_{1}+
\gamma_{a}p^{R}_{a}\cos\theta_{a}\cr 
p^{R}_{a}\sin\theta_{a}\cos\phi_{a}\cr
p^{R}_{a}\sin\theta_{a}\sin\phi_{a}\cr
\gamma^{0}_{a}p^{R}_{a}\cos\theta_{a}+\gamma_{a}E^{R}_{1}\cr}
\end{equation}
and
\begin{equation}
P_{2}=\pmatrix{
\gamma^{0}_{a}E^{R}_{2}-\gamma_{a}p^{R}_{a}\cos\theta_{a}\cr 
-p^{R}_{a}\sin\theta_{a}\cos\phi_{a}\cr
-p^{R}_{a}\sin\theta_{a}\sin\phi_{a}\cr 
-\gamma^{0}_{a}p^{R}_{a}\cos\theta_{a}+\gamma_{a}E^{R}_{2}\cr}
\end{equation}
with
\begin{eqnarray}
E^{R}_{1}&=&\frac{s_a+m_1^2-m_2^2}{2\sqrt{s_{a}}},\nonumber\\
E^{R}_{2}&=&\frac{s_a+m_2^2-m_1^2}{2\sqrt{s_{a}}},\nonumber\\
p^{R}_{a}&=& \frac{1}{2}\,\sqrt{\frac{\lambda(s_{a}, m_{1}^{2},
    m_{2}^{2})}{s_{a}}}.
\end{eqnarray}
The boost-factors are
\begin{eqnarray}
\gamma^{0}_{a}&=&\frac{s+s_{a}-s_{b}}{2\sqrt{ss_{a}}},\nonumber\\
\gamma_{a}&=&\frac{\sqrt{\lambda (s, s_{a}, s_{b})}}{2\sqrt{ss_{a}}}.
\end{eqnarray}
For the other subsystem we have
\begin{equation}
P_3=\pmatrix{\gamma^{0}_{b}E^{R}_{3}-
\gamma_{b}p^{R}_{b}\cos\theta_{b}\cr 
p^{R}_{b}\sin\theta_{b}\cos\phi_{b}\cr
p^{R}_{b}\sin\theta_{b}\sin\phi_{b}\cr
\gamma^{0}_{b}p^{R}_{b}\cos\theta_{b}-\gamma_{b}E^{R}_{3}\cr}
\end{equation}
and
\begin{equation}
P_{4}=\pmatrix{
\gamma^{0}_{b}E^{R}_{4}+\gamma_{b}p^{R}_{b}\cos\theta_{b}\cr 
-p^{R}_{b}\sin\theta_{b}\cos\phi_{b}\cr
-p^{R}_{b}\sin\theta_{b}\sin\phi_{b}\cr 
-\gamma^{0}_{b}p^{R}_{b}\cos\theta_{b}-\gamma_{b}E^{R}_{4}\cr}.
\end{equation}
The energies and boost factors are the same as before with the
substitutions
$a\leftrightarrow b$, $1\to3$ and $2\to4$.
 

\end{document}